\documentclass[fleqn,usenatbib]{mnras}

\usepackage{newtxtext,newtxmath}

\usepackage[T1]{fontenc}

\DeclareRobustCommand{\VAN}[3]{#2}
\let\VANthebibliography\thebibliography
\def\thebibliography{\DeclareRobustCommand{\VAN}[3]{##3}\VANthebibliography}

\usepackage{graphicx}	
\usepackage{amsmath}	
\usepackage{subfig}
\usepackage{mwe}
\usepackage[dvipsnames]{xcolor}
\usepackage[export]{adjustbox}
\usepackage{booktabs}
\usepackage{multirow}

\usepackage{scalerel,tikz}
\usetikzlibrary{svg.path}
\definecolor{orcidlogocol}{HTML}{A6CE39}
\tikzset{orcidlogo/.pic={
 \fill[orcidlogocol] svg{M256,128c0,70.7-57.3,128-128,128C57.3,256,0,198.7,0,128C0,57.3,57.3,0,128,0C198.7,0,256,57.3,256,128z};
 \fill[white] svg{M86.3,186.2H70.9V79.1h15.4v48.4V186.2z}
 svg{M108.9,79.1h41.6c39.6,0,57,28.3,57,53.6c0,27.5-21.5,53.6-56.8,53.6h-41.8V79.1z M124.3,172.4h24.5c34.9,0,42.9-26.5,42.9-39.7c0-21.5-13.7-39.7-43.7-39.7h-23.7V172.4z}
 svg{M88.7,56.8c0,5.5-4.5,10.1-10.1,10.1c-5.6,0-10.1-4.6-10.1-10.1c0-5.6,4.5-10.1,10.1-10.1C84.2,46.7,88.7,51.3,88.7,56.8z};
}}
\newcommand\orcidicon[1]{\href{https://orcid.org/#1}{\mbox{\scalerel*{
\begin{tikzpicture}[yscale=-1,transform shape]
\pic{orcidlogo};
\end{tikzpicture}
}{|}}}}
\newcommand{\al}{\alpha}
\newcommand{\vfrag}{v_{\rm{frag}}}

\title[Can Close-In Exoplanets form by Pebble Accretion?]{Can Close-In Exoplanets form by Pebble Accretion?}

\author[Narayan et al.]{
Jayashree Narayan,$^{\orcidicon{0009-0007-8280-1010}\,1,2}$\thanks{E-mail: \href{mailto:jayashreenarayangs@gmail.com}{jayashreenarayangs@gmail.com}},
Joanna Dr{\k{a}}{\.z}kowska,$^{\orcidicon{0000-0002-9128-0305}\,2}$ and
Vignesh Vaikundaraman$^{\orcidicon{0000-0002-2451-9574}\,2}$
\\
$^{1}$Indian Institute for Science Education and Research, Mohali, Knowledge city, Sector~81, SAS Nagar, Punjab 140306\\
$^{2}$Max Planck Institute for Solar System Research, Justus-von-Liebig-Weg~3, 37077~G{\"o}ttingen, Germany\\
}

\date{Accepted XXX. Received YYY; in original form ZZZ}

\pubyear{\the\year{}}

\begin{document}
\label{firstpage}
\pagerange{\pageref{firstpage}--\pageref{lastpage}}
\maketitle

\begin{abstract}
Pebble accretion is the leading theory for the formation of exoplanets more massive than the Earth. Many parameters influence planet growth in the pebble accretion models. In this paper, we study the influence of pebble fragmentation velocity, turbulence strength, stellar metallicity, stellar mass, and planet location on the growth of planets located within 1~au from their parent stars. Analysing the close-in planets from NASA's Exoplanet Archive, we find that the turbulence strength influences planet growth more than the pebble fragmentation velocity does. Planets orbiting stars with higher metallicity have an overall higher probability of reaching their pebble isolation mass than those orbiting lower metallicity stars, but the impact of metallicity is not as high as that of stellar mass, orbital separation, and most importantly disc turbulence.
\end{abstract}




\section{Introduction}
\label{sec:introduction}

There are two main ways in which planets grow from dust, the classical or planetesimal accretion paradigm and the pebble accretion paradigm. The classical paradigm suggests that solids are quickly and efficiently converted into planetesimals, which continue to grow by what is known as the `runaway growth' \citep{1989Icar...77..330W}. Planetesimals are gravitationally bound objects, and in the runaway growth phase, planetesimal accretion is quick. As the planetesimals increase in size, the eccentricities and inclinations of their orbits also increase and so planet growth then transitions from the runaway growth phase to the `oligarchic growth' phase \citep{1998Icar..131..171K,2010ApJ...714L.103O}. In this phase, the embryo slows its growth down since there is less material for it to accrete from the protoplanetary disc. Towards the end of the oligarchic growth phase, the local feeding zone within the protoplanetary disc becomes devoid of planetesimals. The mass of the planet at this stage is called its pebble isolation mass. If the embryo becomes massive enough and the disc still contains gas, gas accretion onto the core starts \citep{1996Icar..124...62P}. The main drawback with the classical paradigm is that the time taken for the growth to occur is very long, so effective gas accretion cannot occur before disc dispersal, and so giant planets cannot form \citep{2002AJ....123.2862T, 2019A&A...631A..70J}. Furthermore, dust growth to kilometre sizes is hindered by many growth barriers like bouncing, fragmentation, and radial drift \citep{2018SSRv..214...52B}. Due to these drawbacks, the `pebble accretion' model was introduced. Pebble accretion helps giant planet cores to grow more rapidly. 


 Pebble accretion is a pivotal process in the formation of planets, where solid pebbles drifting through a protoplanetary disc interact with planetary cores. In the protoplanetary disc, the gas is pressure-supported, resulting in its slower orbital velocity compared to larger objects. The influence of the gas on the motion of solids varies based on their size; dust moves in tandem with the gas, while the largest planetesimals remain largely unaffected by the gas. Pebbles, however, present an intermediate scenario; aerodynamic drag causes them to settle towards the central plane of the disc and orbit at a sub-Keplerian velocity, leading to radial drift towards the central star \citep{1977MNRAS.180...57W}. These pebbles, affected by gas drag, slow down and become gravitationally bound to larger bodies, eventually spiralling inwards and contributing to growth \citep{2010A&A...520A..43O}. This process can significantly accelerate the growth of planet cores compared to the classical scenario \citep{2012A&A...544A..32L}. For instance, \citet{2012A&A...546A..18M} demonstrated that pebble accretion could quickly build Earth-sized cores in a few thousand years, compared to much longer timescales predicted by traditional models. Most importantly, pebble accretion solves the long-standing challenge of forming Jupiter's core before the gas disc dispersal \citep{2014A&A...572A.107L, 2022MNRAS.510.1298A}.

 There are many parameters which influence planet growth by pebble accretion. Some of these are stellar parameters like the mass, radius and metallicity of the parent star. Some of these are planetary parameters like the mass, radius, and location of the planet. The others are environmental parameters like the mass, size, and temperature of the disc, its turbulence level, fragmentation threshold velocity of dust aggregates, and their density. These parameters may influence planetary growth in their own ways. For example, it is known that protoplanetary discs evolve faster around high-mass (>2 $M_\odot$) stars \citep{2015A&A...576A..52R}, which gives less time for the formation of gas-rich planets to complete. It is also well-known that the higher the metallicity of the parent star, the higher the occurrence of giant planets \citep{2021ApJS..255...14F, 2005A&A...437.1127S}, while the influence of metallicity on the occurrence of smaller planets remains debated \citep{2015AJ....149...14W, 2021AJ....162...69K}. In this paper, we study the importance of disc turbulence, stellar mass and metallicity, planet semi-major axis, and dust fragmentation threshold velocity on the formation of close-in planets. 

 This paper is organized as follows. Section \ref{sec:model} explains the numerical Protoplanetary disc model, the Pebble Accretion model we are working with, the planet growth model, the parameters under study and the pebble flux model. Section \ref{sec:results} gives a description of the data we are using, and discusses the dependence of the model on the parameters under study. It also discusses the effects of varying starting time and analyses the planets that do not get reproduced with our model. Section \ref{sec:discussion} discusses our work and the drawbacks of our model, and finally, in Sect.~\ref{sec:conclusion}, we present our conclusion.

\section{Model}
\label{sec:model}

\subsection{Protoplanetary disc model}
\label{sec:protoplanetary_disc_model}
We consider one-dimensional models of protoplanetary discs surrounding stars of various masses $M_{\star}$. Following \citet{2021ApJ...920L...1M}, we assume that the protoplanetary disc temperature is set to 
\begin{equation} \label{eq:1}
    T = 280\ {\mathrm{K}} \cdot \left( \frac{r}{\mathrm{au}}\right)^{-0.5} \left(\frac{M_{\star}}{M_\odot}\right)^{0.5}, 
\end{equation}
 where $r$ is the radial distance to the central star. We set the initial gas surface density according to the self-similar profile \citep{1974MNRAS.168..603L} 
\begin{equation} \label{eq:2}
    \Sigma_{\rm{gas,0}} = \frac{M_{\rm{disc}}}{2 \pi r_c^2} \left(\frac{r}{r_c} \right)^{-1} \exp{\left( -\frac{r}{r_c} \right) },   
\end{equation}
where $M_{\rm{disc}}$ is the initial disc mass and $r_c$ is the critical radius. The dust surface density is then given by 
\begin{equation} \label{eq:3}
    \Sigma_{\rm{dust,0}} = Z \cdot \Sigma_{\rm{gas,0}},
\end{equation}
 where $Z$ is the dust-to-gas ratio. Since the gas cloud collapses into the star and the disc, the metallicity of the star and the disc should be the same. Thus, we assume that the dust-to-gas ratio is the same as the stellar metallicity. Since we aim to study the influence of a few selected parameters on the growth of planets in a protoplanetary disc, we keep the other parameters fixed. These values are as given in Table~\ref{table:1}.

\begin{table}
\caption{Disc Parameters}\label{table:1} 
\begin{tabular}{lcc}
\hline
Parameter & Value \\
\hline
Disc mass ($M_{\rm{disc}})$      & 0.1 M$_\star$ \\
Planet internal density & 3 g/cm$^{3}$\\
Critical radius ($r_c$) & 30~au \\
\hline
\end{tabular}
\end{table}

 We assume the initial disc mass is 10\% of the stellar mass, roughly corresponding to young, Class 0/I discs \citep{2020A&A...640A..19T}. The dust mass is set by the product of the parent star metallicity, the influence of which we discuss in section \ref{sec:dependence_on_z}, and the total disc mass. For close-in planets, the size of the disc only regulates the speed with which the planet grows, but not the final outcome \citep{2021A&A...647A..15D}, which is why we chose $r_c$ as 30 au for the critical radius.

\subsection{Pebble Accretion Model}
\label{sec:pebble_accretion_model}

 We simulate the formation of a single-planet system through pebble accretion in a protoplanetary disc. The \texttt{pebble predictor}\footnote{\href{https://github.com/astrojoanna/pebble-predictor}{https://github.com/astrojoanna/pebble-predictor}} calculates the time-dependent pebble flux at every location in the disc \citep{2021A&A...647A..15D} and the pebble accretion efficiency is obtained by the numerical fits\footnote{\href{https://staff.fnwi.uva.nl/c.w.ormel/software.html}{https://staff.fnwi.uva.nl/c.w.ormel/software.html}} from \citep{2018A&A...615A.138L} and \citep{2018A&A...615A.178O}. Table \ref{table:1} describes all the initial parameter values we use to grow our planet.

 Pebble accretion efficiency $\epsilon$ is defined as 
\begin{equation} \label{eq:4}
    \epsilon = \frac{\dot{M}_{\textrm{PA}}}{\dot{M}_{\textrm{peb}}},
\end{equation}
where \(\dot{M}_{\textrm{PA}}\) is the pebble accretion rate on the planet and \(\dot{M}_{\textrm{peb}}\) is the mass flux of pebbles through the disc. $\epsilon$ is essentially the probability of a pebble getting accreted onto a planet. The higher the probability, the more efficient the planet's growth is. Note that since the \texttt{pebble predictor} does not calculate the size of the pebbles but only the Stokes number \citep{2021A&A...647A..15D}, the drag regime in which we are working is not relevant.

 Since the accretion efficiency depends on the radius of the planet as the planet grows, on the Stokes number of pebbles that get accreted, and on the height of the pebble layer, both the pebble Stokes number and their settling sensitively depend on the turbulence strength $\al$ and fragmentation threshold $\vfrag$, as discussed in section \ref{sec:dependence_on_alpha_vfrag}.

 From equation \ref{eq:4}, we calculate the mass that gets added onto the planet at every time step $\mathrm{dt}$ as 
\begin{equation}\label{eq:5}
 \text{dM}_{\textrm{PA}} = \epsilon  \dot {M}_{\textrm{peb}} \cdot \mathrm{dt}.
\end{equation}

 We assume that if the planet should grow to its full extent, then it must reach its pebble isolation mass within 3~Myr. In the pebble accretion scenario, the pebble isolation mass is determined by the planet-disc interactions leading to a gap opening in the gas disc which results in a pressure trap arising outside of the planet orbit and cutoff of the pebble flux. Following \citet{2017ASSL..445..197O}, we calculate the pebble pebble isolation mass as
\begin{equation} \label{eq:6}
    M_\text{iso} = 40 M_{\oplus}  \frac{M_*}{M_{\odot}}  \left(\frac{H}{0.05  r}\right)^3,\
\end{equation}
where $H$ is the disc pressure scale height, which is given by 
\begin{equation}\label{eq:7}
    H = \frac{c_s}{\Omega_K},
\end{equation}
 where $c_s$ is the sound speed and $\Omega_K$ is the Keplerian frequency. It is worth noting that the pebble scale height is connected to the gas scale height as follows \citep{2007Icar..192..588Y}:
 \begin{equation}\label{eq:8}
     H_\text{peb} = H\left[\left(1+\frac{\mathrm{St}}{\al}\right) \cdot \left(\frac{1+2\mathrm{St}}{1+\mathrm{St}}\right) \right]^{-1/2},
 \end{equation}
 where $\mathrm{St}$ is the Stokes number of pebbles and $\al$ is the turbulence strength of the disc. The Stokes number is a dimensionless parameter describing the coupling between the pebbles and the gas.

 Pebble accretion can be broadly classified into two regimes: 2D and 3D. Based on the radius of pebble accretion $r_\text{PA}$, which determines the maximum distance from the planetary core within which pebbles are accreted, and the pebble scale height given by Eq.~\ref{eq:8}, the regimes can be clearly defined. In the 2D regime, $r_\text{PA}>H_\text{peb}$ the embryo can accrete from the entire layer of pebbles, whereas in the 3D regime where $r_\text{PA}<H_\text{peb}$, the accretion remains limited to only a fraction of the entire layer of pebbles. Thus accretion in the 2D mode is typically faster \citep{2014A&A...572A.107L, 2016A&A...586A..66V, 2016A&A...596L...3I}. As planetary cores grow larger, their accretion mode transitions from 3D to 2D.

\subsection{Planet Growth Model}
\label{sec:planet_growth_model}
 We consider the growth of the planets to occur in three main stages. The first stage is the growth of dust into pebbles and the formation of planetesimals via the streaming instability. The second is the growth of the planetesimals into planetary cores, and the final stage is the growth of the planetary cores into planets by pebble accretion. We study how the time taken for the first stage, covering the transition from dust to planetesimals (this time is defined as $t_0$) affects planet formation in Sect.~\ref{sec:effect_of_varying_start_time}. Assuming $t_0$ to be 0 initially, meaning planetesimals are formed before the start of our model, our default model considers only stages 2 and 3 for the planet formation. 

 We start with an already existing planetesimal and assume that its initial mass is $10^{-6}M_{\oplus}$, which corresponds to a size of approximately 100~km consistent with the streaming instability formation scenario \citep[see, e.g.,][]{2016ApJ...822...55S} and that the planetesimal accretion dominates until the embryo reaches the classical isolation mass, which is given by
\begin{equation}\label{eq:9}
\begin{split}
    M_\text{iso,plts} & = 2 \pi r \Delta r \Sigma_\text{plts} \\
    & \approx 0.1M_{\oplus} \left(\frac{\Sigma_\text{plts}}{5\ \mathrm{g\ cm}^{-2}}\right)^{3/2} \left(\frac{r}{\mathrm{au}} \right) ^{3} \left(\frac{M_\star}{M_\odot} \right) ^{-1/2},
\end{split}
\end{equation}
where $\Sigma_\text{plts}$ is the surface density of planetesimals and the width of accretion zone is assumed to be ten Hill radii $\Delta r \approx 10r_H$ \citep{1998Icar..131..171K}. The planetesimal growth stage itself proceeds in two phases -- first the runaway growth from the initial planetesimal mass to a transition mass. The transition mass is given by \(M_t = \frac{4\pi}{3}\rho R^3_{\textrm{rg/oli}}\) where $\rho$ is the average density of the planetesimals (see Table \ref{table:1}) and $R_{\textrm{rg/oli}}$ is the radius corresponding to the transition mass  \citep{2010ApJ...714L.103O}, given by
\begin{multline}\label{eq:10}
    R_{\textrm{rg/oli}} = 580 \cdot \left(\frac{\mathrm{C_{\textrm{rg}}}}{0.1}\right) ^{3/7} \left(\frac{R_0}{10\ \mathrm{km}} \right) ^{3/7}\\
    \left(\frac{r}{4\ \mathrm{au}}\right)^{5/7} \left(\frac{\Sigma_\text{plts}}{3\ \mathrm{g\ cm}^2}\right)^{2/7}\ \mathrm{km},
\end{multline}
where $C_{\textrm{rg}} \approx 0.1$ is a numerically determined factor. When the embryo reaches $M_t$, the accretion slows down as it switches to the oligarchic growth regime. In this regime, the rate of mass accretion given by 
\begin{equation} \label{eq:11}
    \dot M_\text{p,oli} = \pi \Omega_K \Sigma_\text{plts} R_p^2.
\end{equation}
This is the second phase of planetesimal growth. This two-phase planetesimal growth is important because its duration determines when the pebble accretion begins. The time needed for the planetesimal to reach the transition mass $M_t$ from the initial planetesimal mass $M_0 = 10^{-6} M_\oplus$ is given by \(N\tau_{\textrm{rg}}\), where
\begin{equation} \label{eq:12}
    N = \log_{2} \left(\frac{M_t}{M_0} \right)
\end{equation}
and 
\begin{equation} \label{eq:13}
    \tau_{\textrm{rg}} \approx C_{\textrm{rg}} \frac{\rho R_0}{\Omega_K \Sigma_\text{plts}},
\end{equation}
with $\tau_{\textrm{rg}}$ being the runaway growth timescale and $R_0$ being the radius of the accreted planetesimals \citep{2023ASPC..534..717D}. The time needed for the planetesimal to grow from $M_t$ to the classical isolation mass equation \ref{eq:9} is given by integrating equation \ref{eq:11}. Summing up the two contributions (from the 2 phases of planetesimal growth), we obtain the time at which pebble accretion starts. 

 The growth of the planets is modelled at the same locations relative to the parent star they exist in the present day. These locations are obtained from the exoplanetary archive. They are all within 1 au, which is why we call them `close-in' planets. Since these planets are located so close to their parent stars, we have ignored any possibilities of planet migration. We assume that the embryos are in circular orbits and with zero inclination.

\subsection{Parameters under study}
\label{sec:params_under_study}
We study the influence of a total of three parameters. Two of these are disc parameters and the other is a stellar parameter. The disc parameters are the turbulence coefficient $\al$ and the pebble fragmentation velocity $\vfrag$. The stellar parameter is the stellar metallicity, which is also called the dust-to-gas ratio $Z$. They are defined as follows:

 1.~Turbulence coefficient ($\al$): This parameter (derived from the Shakura-Sunyaev $\al$ parameter, \citealt{1973A&A....24..337S}) measures the efficiency of angular momentum transport due to turbulence. It also defines turbulence strength, which translates into collision speeds between pebbles as well as the settling of pebbles to the midplane. 

 2.~Fragmentation velocity ($\vfrag$): When pebbles collide with relative velocities larger than $\vfrag$, then the collisions lead to fragmentation rather than growth \citep{2010A&A...513A..56G}. Thus, together with $\al$, this parameter regulates the maximum Stokes number of pebbles.

 3.~Metallicity of the gas cloud ($Z$): This is the ratio of metals to the hydrogen-helium gas in the parent star. Note that here, `metals' are defined as every element heavier than helium. We assume that this is equivalent to the initial dust-to-gas ratio in the disc and the current metallicity of the planet-hosting star. 

 The pebble flux  $\dot M_\text{peb}$ in Equation \ref{eq:5} depends on the three parameters described above. Since pebble accretion can occur as long as the disc lives, we take the end of pebble accretion to be similar to that of the disc lifetime, which is around 3~Myr \citep{2009AIPC.1158....3M, 2015A&A...576A..52R}. If we were to change any of the above parameters, the mass that gets added onto the planet at every time step would change significantly, leading to a noticeable change in the total mass of the planet from t=0 to t=3~Myr. Consequently, these parameters can decide whether or not the planet reaches its pebble isolation mass from the planetary core within 3~Myr, thus drastically changing the outcome of our simulations.

\subsection{Pebble Flux Model}
\label{sec:pebble_flux_model}
The value of metallicity  $Z$ is obtained directly from the exoplanetary archive, the turbulence strength $\al$ and fragmentation velocity $\vfrag$ are used to calculate the pebble Stokes number and flux in the \texttt{pebble predictor}. By utilizing the initial gas and dust surface densities, temperature, turbulence strength, and fragmentation threshold velocity, the code predicts properties like the flux-averaged Stokes number and pebble flux across the disc. It assumes that the dust grains are initially micron-sized and grow by collisions with other grains. The dust growth timescale is given by
\begin{equation} \label{eq:14}
    \tau_{\mathrm{growth}} = \frac{\Sigma_{\mathrm{gas},0}}{\Sigma_{\mathrm{dust},0}} \cdot \Omega_K ^{-1} \cdot \left( \frac{\al}{10^{-4}} \right)^{-1/3} \cdot \left(\frac{r}{\mathrm{au}}\right) ^{1/3}.
\end{equation}
where $\Sigma_{\mathrm{dust},0}$ and $\Sigma_{\mathrm{gas},0}$ are defined in Eqs~\ref{eq:2} and \ref{eq:3}. This growth continues until either fragmentation or the radial barrier stops it. Since we are focused on close-in planets, the fragmentation barrier is relevant to our models more than the radial barrier \citep[see, e.g.,][]{2012A&A...539A.148B}. The maximum Stokes number for dust aggregates is then determined as
\begin{equation}\label{eq:15}
    \mathrm{St}_{\mathrm{frag}} = \mathrm{f_f}\frac{\vfrag^2}{3 \al c_s^2},
\end{equation}
where $\mathrm{f_f}=0.37$ is a numerically determined parameter.

The pebble flux can be calculated as 
\begin{equation}\label{eq:16}
\dot{M}_\text{peb} = 2 \pi r v_r \Sigma_{\mathrm{dust}},
\end{equation}
where $\Sigma_{\mathrm{dust}}$ is the surface density of dust in the disc at a given point of time, and $v_r$ is the radial drift velocity of pebbles calculated as 
\begin{equation} \label{eq:17}
v_{r} = \frac{ 2 v_{\eta} \mathrm{St}}{1+\mathrm{St}^2} + \frac{v_{\mathrm{gas}}}{1+\mathrm{St}^2},
\end{equation}
where $v_{\eta}$ is the maximum drift speed, and $v_{\mathrm{gas}}$ is the speed of accretion flow of gas. Since the pebble flux depends on the pebble Stokes number through the radial speed of dust, the parameters we study $\al$ and $\vfrag$ influence the growth of planets both through the pebble accretion efficiency $\epsilon$ and pebble flux in Eq.~\ref{eq:5}. For further details on how the pebble predictor works, please refer to \citet{2021A&A...647A..15D}.

\section{Results}
\label{sec:results}

To understand the dependence of $\al$ and $\vfrag$, we use a range of values of both parameters and monitor the growth of each planet under various possible combinations of the 2 parameters. We varied the values of $\al$ between $10^{-5}$ and $10^{-2}$, a range typically used in planet formation models, and the values of $\vfrag$ between 1~m/s and 10~m/s. We created a linear grid spaced every 1~m/s for the $\vfrag$ values and a logarithmic grid for the $\al$ values ($10^{-5}, 10^{-4}, 10^{-3}, 10^{-2}$). This makes 10 values for $\vfrag$ and 4 values for $\al$, thus giving us $4\times 10 =40$ models. We will run our dataset of 862 planets through each of these models to see the influence of $\al$ and $\vfrag$ on their growth. We describe the details of our dataset in the next section to understand better the types of planets we are working with. To see the influence of metallicity on planet growth, we used the values of metallicity for each planetary system from the exoplanetary archive itself. 

\subsection{Description of the data}
\label{sec:description_of_data}

 In this paper, we study the influence of $\al$, $\vfrag$ and $Z$ on planet growth by pebble accretion for close-in planets. Other parameters like the mass of the parent star $M_{\star}$, the location of the planet (relative to the star), and the radius of the star are obtained from the \href{https://exoplanetarchive.ipac.caltech.edu/cgi-bin/TblView/nph-tblView?app=ExoTbls&config=PS}{NASA exoplanet archive}. Many of these parameters have not been measured for all of the planets in the database. To properly study these parameters, we filter out the planets that don't have proper documentation by applying the `not null' keyword in the search bars of all the relevant columns\footnote{Relevant columns are planet name,	discovery year,	semi-major axis (au), planet radius ($R_ \oplus$),	planet mass (earth mass), density ($g/cm^3$), radius of star ($R _\odot$), mass of star ($M_ \odot$), stellar metallicity (Z = log(k)*$Z_\odot$), Date of publication that released each row of data, and all the errors for each column.}. Once these conditions are lain, we have $\approx$1300 rows. Since many publications have described the same planet, several duplicate rows exist. We retain only the rows with data from the latest publication. Upon applying the maximum distance of the semi-major axis as 1~au and the above condition, we get a total of 862 planets to study\footnote{As of June 2024.}.

\begin{figure}
    \centering
    \includegraphics[width=1\linewidth]{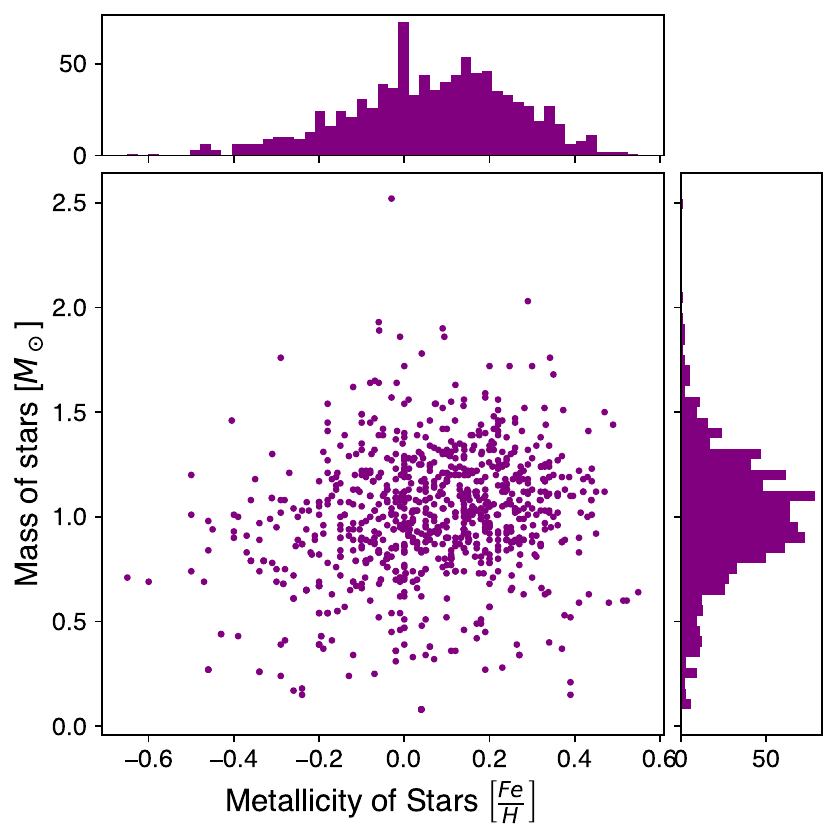}
    \caption{Mass of the planet host stars from our dataset against their metallicities. There is no correlation between the mass of the host starts and their metallicities.}
    \label{fig:1}
\end{figure}

\begin{figure}
    \centering
    \includegraphics[width=1\linewidth]{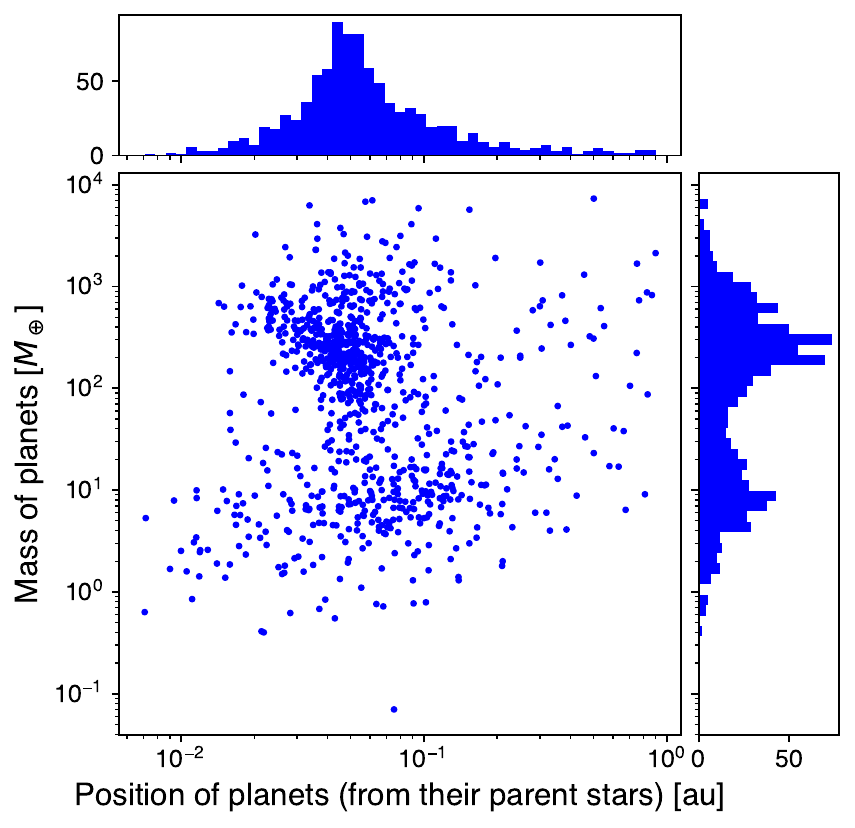}
    \caption{Mass of planets studied here against their orbital distance from the parent star. Most of the planets have a mass between $10^2M_\oplus$ and $10^3M_\oplus$, and most of them exist at around 0.05 au from their host stars.}
    \label{fig:2}
\end{figure}

 Figures \ref{fig:1} and \ref{fig:2} present the data obtained from the NASA Exoplanet Archive. As we can see from Fig.~\ref{fig:1}, the mass of the planet-hosting stars ranges from 0.08~$M_\odot$ to 2.52~$M_\odot$ with a standard deviation of around 0.3~$M_\odot$ and the metallicities range from -0.65 to 0.55 with a standard deviation of around 0.2. From the graph, we can say that there is no particular correlation between these two parameters. From Fig.~\ref{fig:2}, we can see that most of our planets are between $10^{-1}M_\oplus$ and $10^4 M_\oplus$, and they mostly exist at around 0.04 au from their host stars. However, these numbers are impacted by the detection biases - it is easier to detect larger planets (by size). Removing the detection biases revealed that super-Earths and sub-Neptunes are likely the most common types of planets in our Galaxy \citep{2022BAAS...54e2101M}. 
 
\begin{figure}
    \centering
    \includegraphics[width=\linewidth]{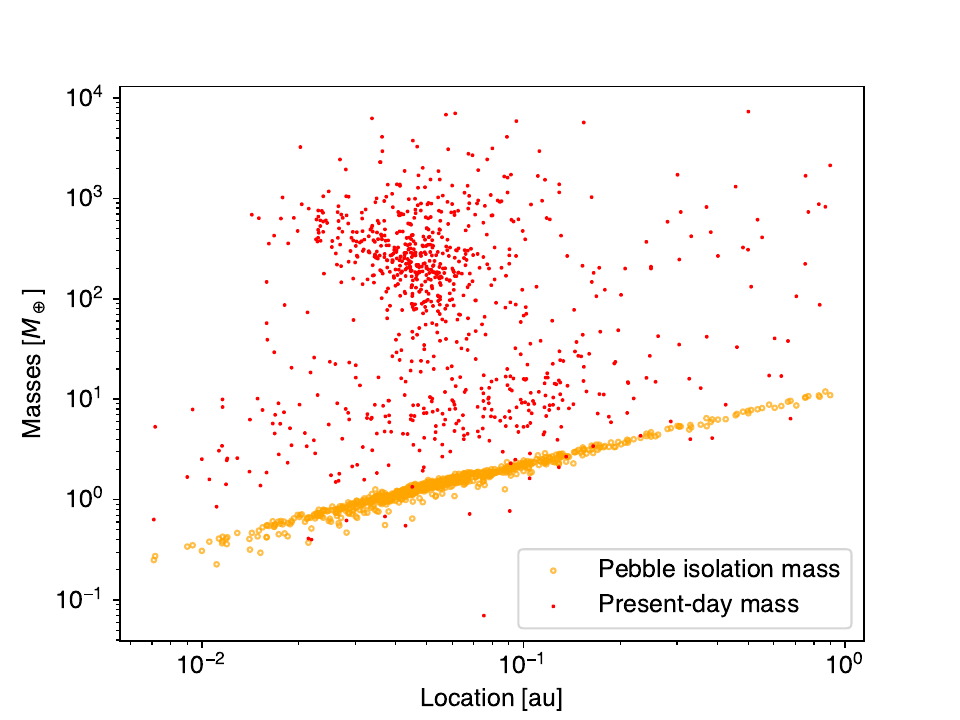}
    \caption{The masses of the planets from the exoplanetary archive are shown in red and the pebble isolation masses calculated for the corresponding disc parameters are shown in yellow.} 
    \label{fig:isol_mass_and_pl_mass}
\end{figure}

\subsection{Dependence on fragmentation threshold and turbulence strength}
\label{sec:dependence_on_alpha_vfrag}

\begin{figure}
    \centering
    \includegraphics[width=\linewidth]{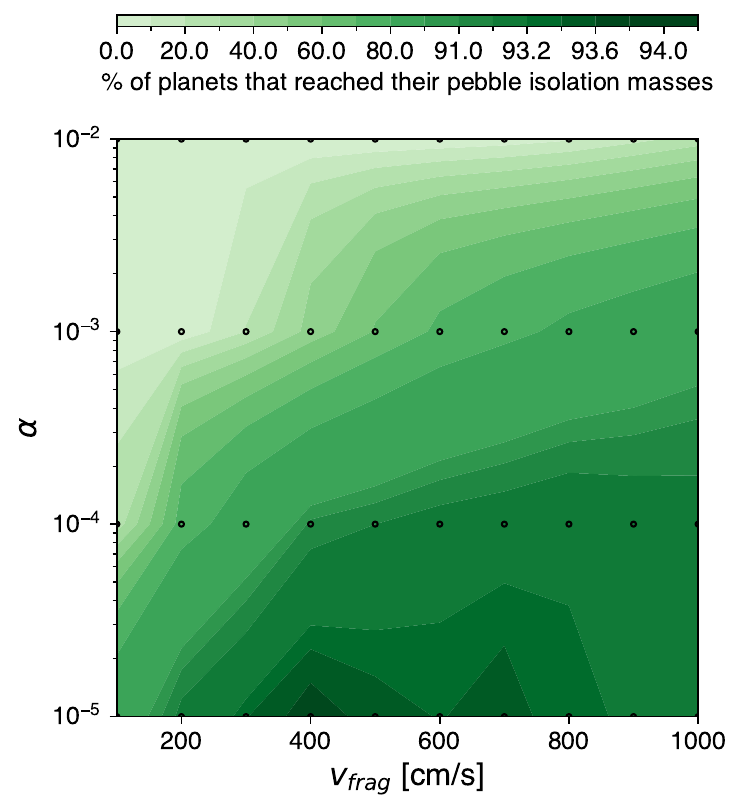}
    \caption{Percentage of planets that reached their pebble isolation mass as a function of pebble fragmentation threshold $\vfrag$ and turbulence strength $\al$. Note that the model with the maximum percentage of planets reaching their pebble isolation masses is at $\al=10^{-5}$ and $\vfrag=4$~m/s at $\approx 94\%$}  
    \label{fig:3}
\end{figure}

\begin{table*}
\caption{The number of planets that reached the pebble isolation mass, for all the 40 models under consideration. The most successful model is highlighted in red.}
\begin{tabular}{c|cccccccccc}
\toprule
$\al $ / $\vfrag$ [m/s]& 1 & 2 & 3 & 4 & 5 & 6 & 7 & 8 & 9 & 10 \\
\midrule
$10^{-5}$ & 756 & 797 & 805 & \textcolor{red}{808} & 806 & 805 & 806 & 804 & 803 & 801  \\
$10^{-4}$ & 209 & 646 & 743 & 787 & 793 & 798 & 800 & 802 & 799 & 797  \\
$10^{-3}$ & 0   & 18  & 174 & 377 & 524 & 622 & 669 & 707 & 730 & 752  \\
$10^{-2}$ & 0   & 0   & 0   & 0   & 4   & 15  & 35  & 74  & 142 & 212 \\
\bottomrule
\end{tabular}
\label{table:no._of_isolated_planets}
\end{table*}

We want to find the number of planets (from our dataset of 862) that reach their pebble isolation mass in each of the 40 models. If the planetary embryo's mass reaches the pebble isolation mass before 3~Myr, then we count the model as successful. In this paper, we do not consider the gas accretion stage of the planets as they grow. As we show in Fig.~\ref{fig:isol_mass_and_pl_mass}, most of the planet masses in the exoplanetary archive have a mass larger than that of the pebble isolation mass our calculation predicts. This is because these planets undergo gas accretion after reaching the pebble isolation mass. In very few cases, the exoplanet mass is lower than the predicted pebble isolation mass. This may be caused by early disk dispersal, pebble flux reduction due to other planets in the same system, or other factors that our model does not include. This is further discussed in Sect.~\ref{sec:discussion}.

Table \ref{table:no._of_isolated_planets} presents the results we obtained and their graphic representation is provided in Fig.~\ref{fig:3}, where the open circles represent the 40 models we evaluated, and the colours correspond to the percentage of planets that reached their pebble isolation mass for that particular model. As we can see from Fig.~\ref{fig:3}, the higher the turbulence in the disc, the harder it is for planets to reach their pebble isolation masses for any value of the fragmentation velocity. But when we consider weakly turbulent discs, even low fragmentation speeds allow for planets to grow. The higher values of $\al$ prevent planet growth for the reason that the pebble growth stops at smaller sizes (see Eq.~\ref{eq:10}), which leads to lower pebble flux (see Eq.~\ref{eq:11}), and the pebble settling to the midplane is made harder (see Eq.~\ref{eq:8}), which means that the pebble accretion rate is lower than for disc with lower $\al$. Similarly, the higher the value of $\vfrag$, the easier the planet growth becomes because pebbles grow to larger sizes, drift faster, and settle better. We can also notice that at larger values of $\vfrag$, for $\al=10^{-4}, 10^{-5}$, there is a dip in the number of planets that reach their pebble isolation mass as $\al$ reduces. This shows that value of $\vfrag$ that are too high, can make it harder for planets to reach their pebble isolation masses. This is because the pebble flux duration is shorter for the higher $\vfrag$ cases and so a fraction of the pebbles is gone before the embryo is formed out of planetesimals.   

\subsection{Dependence on Stellar Metallicity}
\label{sec:dependence_on_z}

\begin{figure*}
\begin{tabular}{ccc}
\includegraphics[width = 0.4\linewidth]{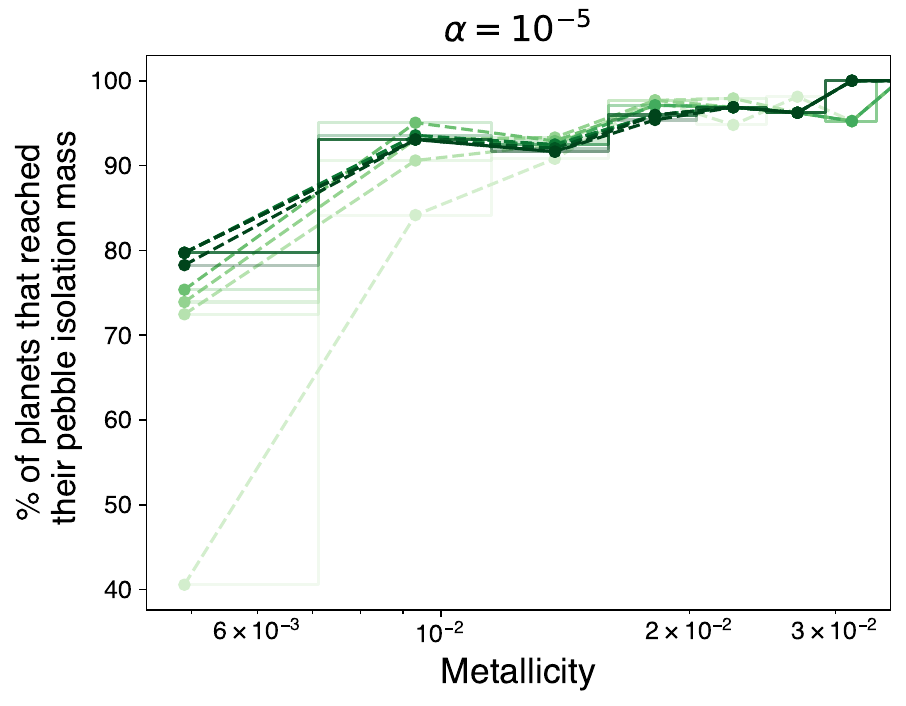} &
\includegraphics[width = 0.4\linewidth]{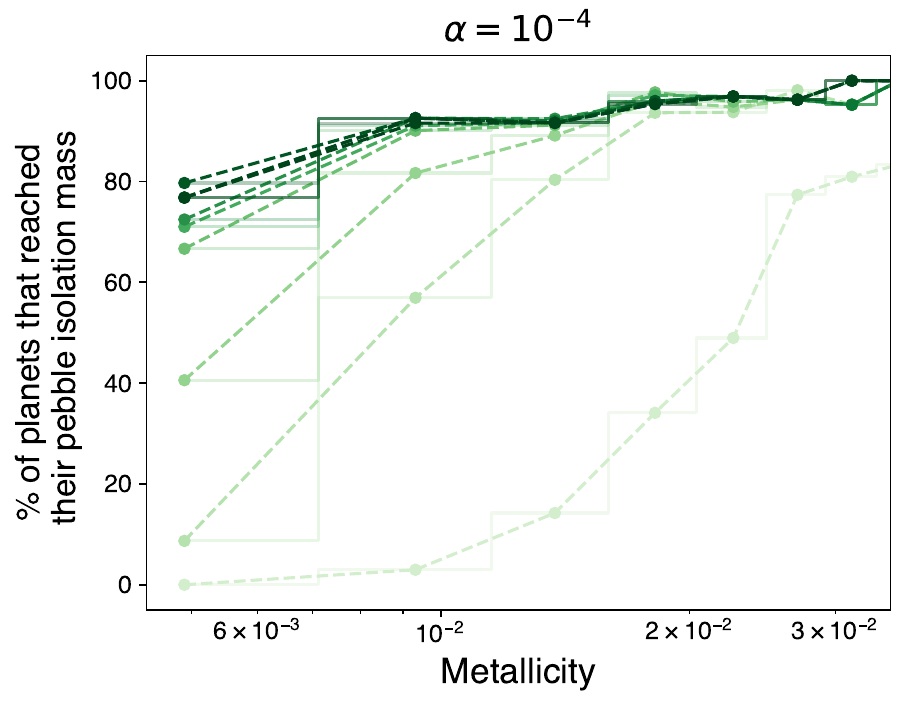} &
\multirow{1}{*}[5.0cm]{\includegraphics[height = 10.5cm]{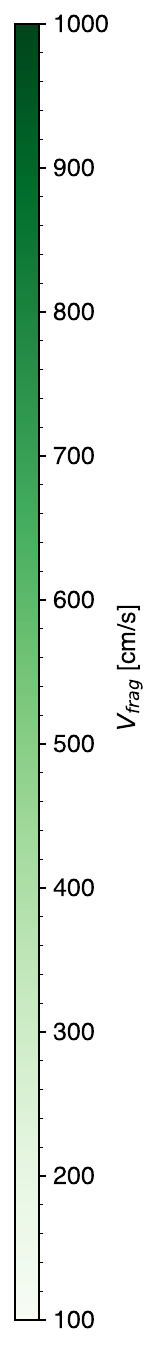}} \\
\includegraphics[width = 0.4\linewidth]{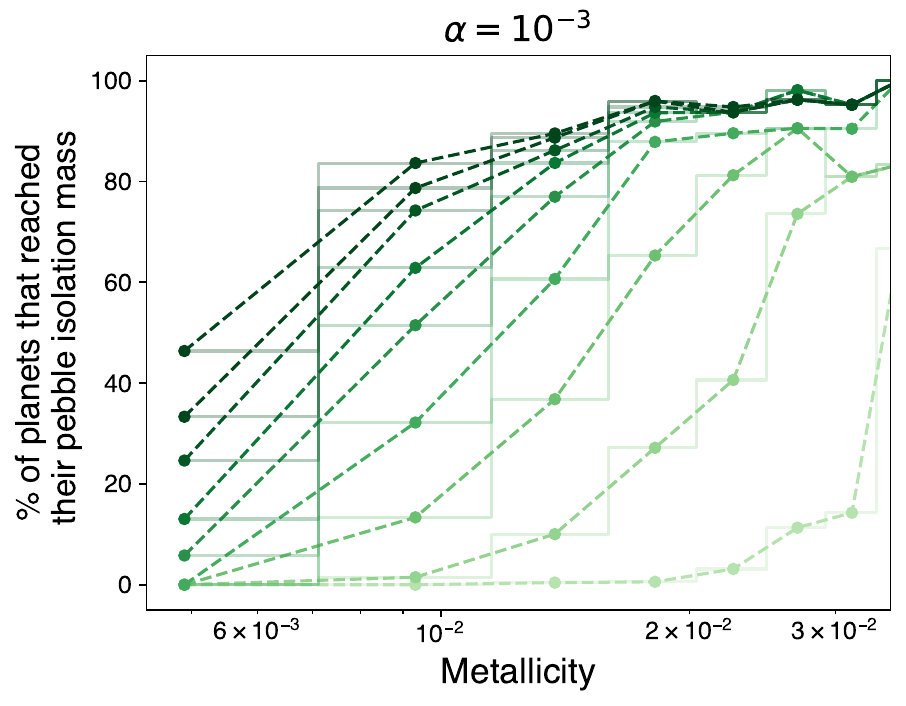}&
\includegraphics[width = 0.4\linewidth]{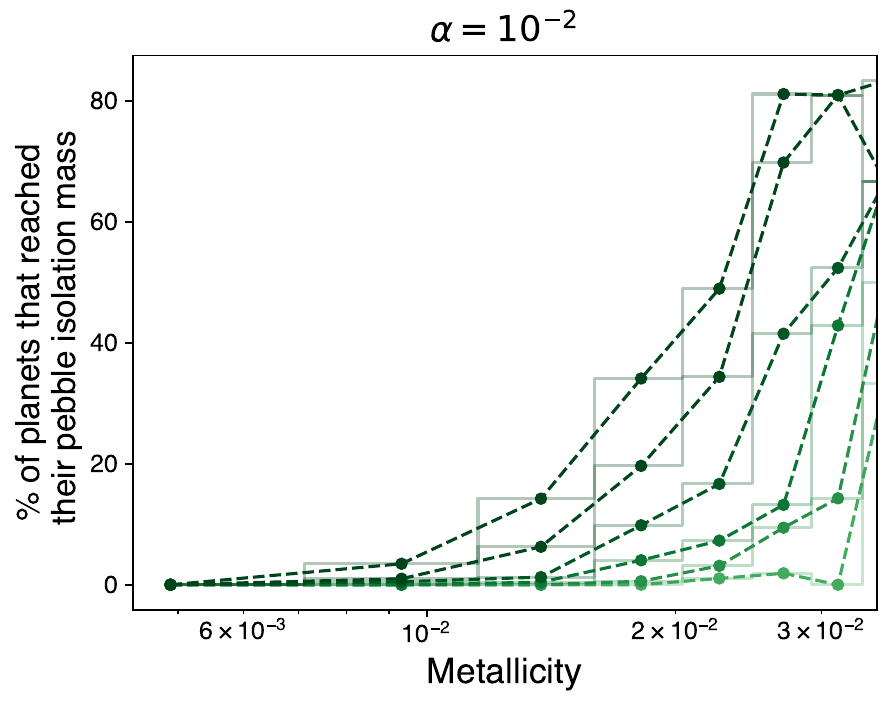}\\
\end{tabular}
\caption{Percentage of planets that reached their pebble isolation mass against the stellar metallicity. The panels correspond to different disk turbulence levels as indicated by their title and the line colours correspond to the assumed fragmentation speed.}
\label{fig:metallicity}
\end{figure*}

 Since we want to study the influence of metallicity on the growth of planets, we need to keep the other parameters constant. This means that we need to look at the metallicity variations of all 862 planets within each of the 40 models. We did this by binning the metallicities of the stars for each model. Then, within each bin, we check what percentage of planets reached their pebble isolation mass. We then plot the percentage of planets that reach their pebble isolation mass in a particular metallicity bin as a function of the metallicity (dimensionless here)\footnote{Here, the metallicity values are converted from the $\left[\frac{\text{Fe}}{\text{H}}\right]$ values displayed in Fig.~\ref{fig:1} to dimensionless metallicities.} of their host stars. Since we are assuming that the metallicity of the gas cloud, the parent star, and the protoplanetary disc are identical, the higher the stellar metallicity, the more pebbles are present in the disc, which makes it more likely for a planet to grow.

 The results of the metallicity analysis are presented in Fig.~\ref{fig:metallicity}. Each panel corresponds to one value of $\al$, and the lines within each graph correspond to different values of $\vfrag$ whose values are given by the colorbar in $\mathrm{cm/s}$. Some of the graphs have a lower number of lines than others -- this is because at $\al = 10^{-2}$ and $10^{-3}$, for some of the models, no planets reach their pebble isolation mass. Although there are dips in the percentage of planets reaching their pebble isolation masses for certain metallicity values, there is an overall rising trend. We can attribute the dips to a statistical issue since we are working with a rather low number of planets.


\subsection{Effects of Varying the Starting Time}
\label{sec:effect_of_varying_start_time}
\begin{figure}
    \begin{tabular}{cc}
    \includegraphics[width = 0.8\linewidth]{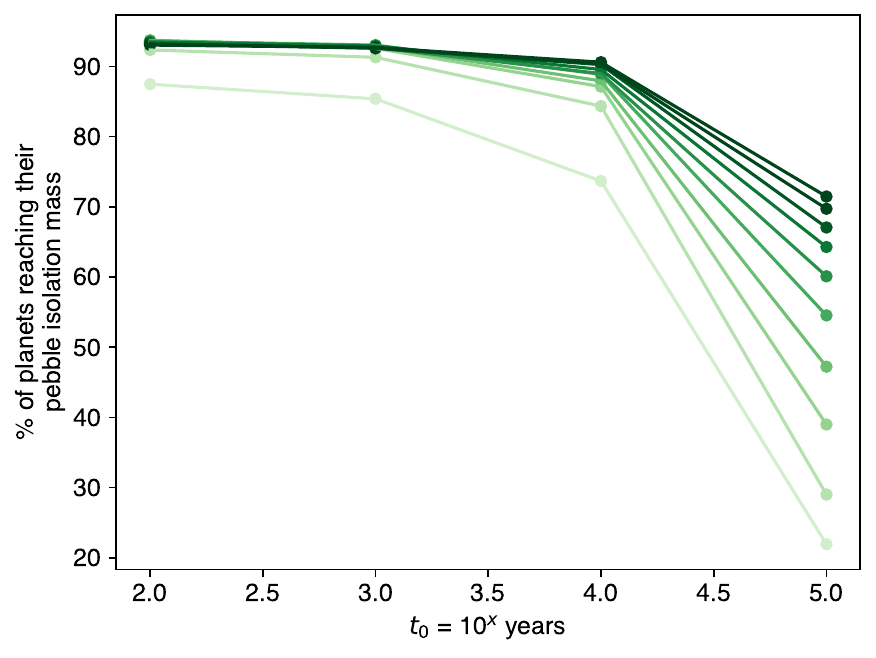} &
    \multirow{1}{*}[4.95cm]{\includegraphics[height = 4.7cm]{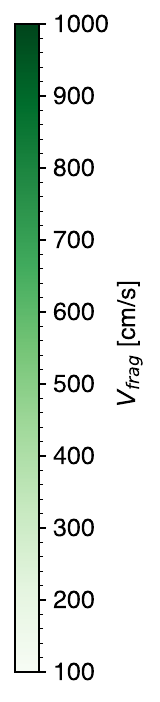}}\\
    \end{tabular}
    \caption{Variation of the percentage of planets that reached their pebble isolation mass at $\al = 10^{-5}$ and for the various values of $\vfrag$ depending on the assumed time of planetesimal formation $t_0$.}  
    \label{fig:5}
\end{figure}

The time taken for dust particles to grow into pebbles and form planetesimals is defined by $t_0$. When $t_0$ is taken to be 0~years, corresponding to planetesimal existing at the start of our model, for example formed during the disc buildup stage \citep{2018A&A...614A..62D}, we find that at $\al = 10^{-5}$ and $\vfrag$ = 4~m/s the largest percentage of planets ($\sim$94\%) reached their pebble isolation mass. We varied the value of $t_0$ to find its effects on the percentage of planets that reach their pebble isolation mass. Figure~\ref{fig:5} shows the result of this study for the constant value of $\al=10^{-5}$ and varying value of the fragmentation speed $\vfrag$. We find that the time taken to form planetesimals drastically changes the success rate of our models. The later we introduce the planetesimals, and thus planetary cores, the fewer planets reach their pebble isolation mass within the disc lifetime. This is because the longer it takes for the core to form, the more pebbles are lost via the radial drift, and the mass budget available for growing the core shrinks. If planetesimals only form after 10$^5$ years (as, e.g., in the models presented by \citealt{2016A&A...594A.105D}), less than 75\% of close-in planets can be reproduced by the pebble accretion model assuming $\vfrag = 10$~m/s. Thus, adding a self-consistent model of planetesimal formation would be necessary to improve our model in the future. 

\subsection{Planets not reproduced with our model}
\label{sec:planets_not_reproduced}

 There is no model out of the 40 we have considered that allows for 100\% of the planets to reach their pebble isolation mass. Of all the models, the highest percentage of planets, around 94$\%$, grow to their maximum mass at $\al = 10^{-5}$ and $\vfrag= 4$~m/s. It is reasonable to ask the question of why this is the case and what are the properties of planets whose growth cannot be reproduced with the pebble accretion model within the parameter space we studied.

\begin{figure*}
    \centering
    \includegraphics[width=\linewidth]{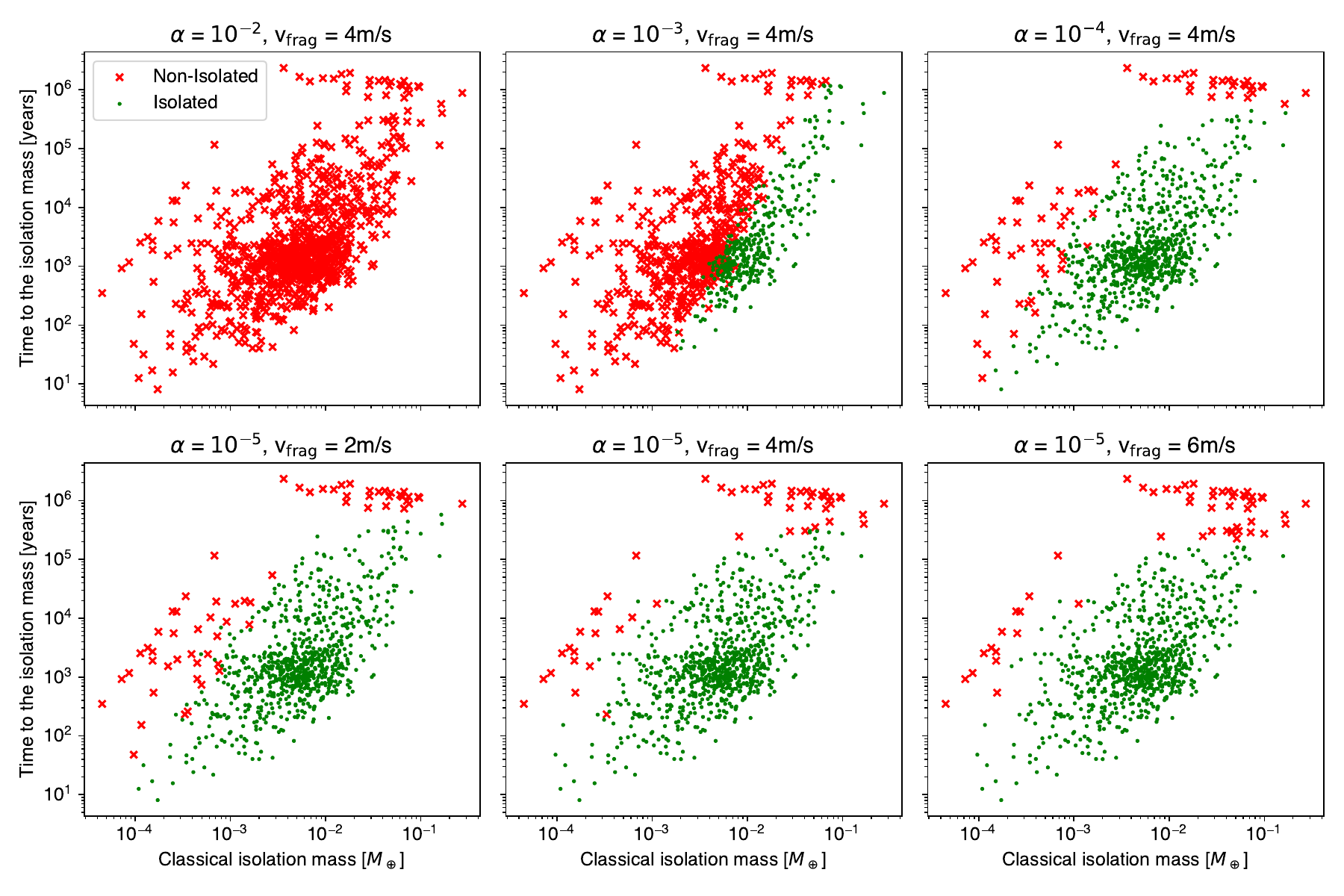}
    \caption{Time taken for the planets to reach the classical isolation mass as a function of the classical isolation mass for the different models. The green dots correspond to models in which the planetary embryo reached the pebble isolation mass, and the red crosses correspond to models in which the planet did not reach the pebble isolation mass. The turbulence strength $\al$ and the fragmentation speed of each set of models are specified in each panel's title.} 
    \label{fig:7}
\end{figure*}

 To answer this question, we studied the planets that did not reach the pebble isolation mass in the above-mentioned (94\%) model and a few other models. Upon analyzing the basic characteristics of these planets (as described in section \ref{sec:description_of_data}), we did not see any direct correlation between the planets that did not reach their pebble isolation mass and their respective characteristics. We noticed that planets around the lowest mass stars in our sample are particularly bad at reaching the pebble isolation mass, but the higher stellar mass does not guarantee a higher success rate for our models. Upon further analysis, we concluded that the planets that fail to reach pebble isolation mass are those that either have low classical isolation mass (below 0.001$~M_{\oplus}$) or reach the classical isolation mass (which we take as transition to pebble accretion) too late, later than about $2\cdot10^5$~years, when most of the pebbles are lost. The first case is more likely to occur in discs surrounding low-mass stars (since we assume the initial disc-to-star mass ratio is constant, the low-mass stars have smaller mass budget available for planet formation), in discs surrounding low-metallicity stars, and for very close-in planets since the classical isolation mass scales with distance (see Eq.~\ref{eq:9}). The second case is more likely to happen for low-mass discs (again disfavouring planet formation around the low-mass stars) and planets further out from their stars, where planetesimal accretion is slower (see Eq.~\ref{eq:11}). This issue might be partially mitigated if we considered concurrent planetesimal and pebble accretion, as discussed further in Sect.~\ref{sec:discussion}. We plotted the time taken for the planets to reach their classical isolation mass against the classical isolation mass itself, marking those that did not reach their pebble isolation mass with a red `x' in Fig.~\ref{fig:7}.

\section{Discussion}
\label{sec:discussion}

Our results suggest that pebble accretion is a generally viable model for forming close-in exoplanets, consistent with earlier works (see, e.g. \citealt{2019A&A...627A..83L, 2021A&A...650A.152I, 2021A&A...647A..15D}). We showed that the main factors influencing the success of the model are higher stellar mass (and thus higher disc mass) and a small distance to the central star. Metallicity seems to be influencing the results to a smaller degree for the planets in our sample (see Fig.~\ref{fig:metallicity}), consistent with the reported weak or even no correlation between super-Earth-sized exoplanets and their host star metallicity \citep{2015AJ....149...14W, 2016AJ....152..187M, 2021AJ....162...69K}. Our work is however impacted by several limitations, as we discuss in this section.

Models of protoplanetary discs suggest that the innermost regions of the disc are fully turbulent \citep[see, e.g.,][]{2017ApJ...835..230F}, with the turbulence level characterized by $\al$ in the range of $10^{-2}$. Thus, the $\al$ value of $10^{-5}$ we find the most favourable for planet growth, or the $\al=10^{-4}$ which still guarantees high success level, may not be reasonable assumptions for the close-in planets. One solution to this problem could be that planets grow mostly in the dead zone and migrate inwards only after reaching significant fraction of their mass in the more quiescent part of the disk. 

We neglect migration of the planetary core through the disc in our models. The same assumption was also made in \cite{2021ApJ...920L...1M} who studied the correlation between stellar mass and types of planets that can be formed by pebble accretion. This may be justified as, in general, the protoplanet formation is expected to proceed the fastest at the inner edge of the disc \citep{2002ApJ...581..666K}. The location of the inner edge of the disc is time-dependent and moves outward as the disc ages. Depending on the timescale of this recession, there are two possibilities. If the timescale for recession is longer than the timescale for planet growth, then the planets can start forming in an inside-out manner, one after another \citep{2014APJ, 2018APJ}. If the opposite is true, then the growth can be truncated by acting as a migration trap, so the planet's location should not change significantly during the disc lifetime \citep{2019A&A...630A.147F}.  

In this paper, we have assumed that the pebble isolation mass of the planets is related only to the mass of the parent star and the disc aspect ratio following the work of \citet{2017ASSL..445..197O}. However, \citet{2018A&A...615A.110A} and \citet{2018A&A...612A..30B} have shown that pebble isolation mass may increase in more turbulent discs. Similarly, \citet{2020A&A...638A..97Z} has shown that pebble isolation mass decreases when we take a realistic equation of state and radiative cooling is assumed to be present.

In our models we assume that planet growth occurs in stages: it proceeds by planetesimal accretion from the initial 100 km-sized planetesimals to the classical isolation mass and then it switches to pebble accretion after that. In reality, pebble accretion can also start earlier, contemporarily with planetesimal accretion \citep[see, e.g.,][]{2022A&A...668A.170L}. This could possibly help us obtain a higher success rate in the models where the core currently forms too late (see Sect.~\ref{sec:planets_not_reproduced}). This would however require more complex modeling where initially inefficient pebble accretion is anyway included in the integration to properly capture its onset.

In our study of the dependence of metallicity on planet growth by pebble accretion, we assume that the metallicity and the dust-to-gas ratio are the same, but this is not necessarily true. \cite{2023A&A...678A..74N} suggests that the types of elements and the quantity of those elements in the disc, depends on the abundance and temperature of the disc itself.

We assume that if the planetary core reaches the pebble isolation mass, we can count the model as successful despite the detected planet mass being higher because the exoplanets acquired a gaseous envelope. One can ask if pebble isolation masses in the inner disc translate into the possibility of accreting any gas. Runaway gas is excluded as the gas is too hot, so giant planets would need to form in the outer disc and then migrate inward, which we do not include in our model. However, we need to note that there the close-in giant planets do not represent the common outcome of planet formation. 

In our models, we keep the planet bulk density constant at $3$~g/cm$^3$ when we do the mass or radius calculations. However, realistically, this density would change as the planet grows and goes through complicated internal evolution and accretes gaseous envelope \citep[see, e.g.,][]{2023ASPC..534..907L}. Keeping the internal density constant introduces an error in our calculations which is difficult to quantify as the accretion rate depends on the embryo radius both in the planetesimal accretion and in the pebble accretion stage.

Our model does not include the possibility of other planets present in the same protoplanetary discs of the planetary systems we are studying. Such planets can influence the inner planets' growth. They can block the flux of pebbles, thus leaving little to nothing for the inner planets to feed off of \citep{2019A&A...627A..83L, 2021ApJ...920L...1M}. Thus, our pebble flux may be overestimated.

By choosing specific values for both $\al$ and $\vfrag$, we investigate the impact of individual parameters. Although this is a simpler approach than the population synthesis approach, including drawing random parameters from a distribution, our findings suggest that no single parameter set guarantees the success of all models, leading us to propose that, in reality, these parameters likely vary from disk to disk or even within one disk.

In this paper, we have discussed the best values of $\al$ and $\vfrag$ for planet formation by pebble accretion, and in doing so, we have not considered any constraints on turbulence in protoplanetary discs. For example, \citet{2023NewAR..9601674R} reviewed the different techniques that have been devised to observationally measure the level of turbulence in discs. In their Table 2, the authors have shown the existing constraints from the radial width of the gas substructures where the values of $\al$ vary from $10^{-4}$ to $10^{-2}$, so within the levels we covered in our parameter study, although as we discussed above, the lower values may not correspond to the inner disk part which we are interested in.  \citet{2024A&A...682A..32J} and \citet{2024NatAs...8.1148U} discussed pebble size measurements in the observed discs. Both of these works suggest rather fragile dust grains, with the fragmentation speeds closer to the lowest value we covered of $\vfrag=1$ m/s, consistent with the laboratory experiments with silica dust \citep{2010A&A...513A..56G}. Our work shows that the combination of high turbulence with low fragmentation speed is particularly disadvantageous to growing planets by pebble accretion. Thus, there must be at least a region in the disk where sticking is aided, for example, by organic mantles as suggested by \citet{2019ApJ...877..128H}.

Furthermore, we notice the inconsistency between the models that provide the highest success in forming planets and the observed lifetimes of protoplanetary disks. In the models that assume high fragmentation speed and low turbulence, the pebbles are typically gone in $\sim 2 \cdot 10^5$ years, whereas the observed protoplanetary disks lose pebbles on Myrs timescales \citep{2020ARA&A..58..483A}. This problem has been noticed in the disc population study by \citet{2022A&A...661A..66Z}, who suggested that disc substructures that slow down the inward drift must be common. A different approach to this problem was presented by \citet{2023A&A...673A.139A} who showed that the observed disc lifetimes are compatible with radial drift if the disc formation and initial viscous spreading are considered and the turbulence level is relatively high ($\al=10^{-3}$ to $10^{-2}$) and the fragmentation speed is low ($\vfrag=1$~m/s). This discrepancy between parameters allowing to explain the observed properties of discs and efficient planet formation suggested by the exoplanet detections requires further studies. In fact, many planetesimal formation models rely on the process of pebble drift to supply solid material for growth (e.g., \citealt{2023ASPC..534..717D}). Since the rate at which pebbles drift inward influences the accumulation of material in protoplanetary disks, a valuable follow-up study could investigate models where the planetesimal formation time ($t_0$) is directly tied to the pebble drift rate. This would provide deeper insights into how variations in drift efficiency impact the timing and efficiency of planetesimal formation, potentially refining our understanding of early planet formation processes.

\section{Conclusions}
\label{sec:conclusion}
In this work, we have studied the possibility of forming known close-in exoplanets in situ by accretion of pebbles. We selected 862 exoplanets for which parameters needed to build realistic disc models are known well enough. We considered close-in planetary core formation in the classical planetesimal-driven paradigm and we used a state-of-the-art model for dust evolution to calculate pebble flux and pebble Stokes number to model planetary core growth by pebble accretion. We investigated range values of the two disc parameters influencing the pebble evolution that are hard to constrain otherwise: the turbulence strength $\al$ and fragmentation speed $\vfrag$. We found that pebble accretion can successfully reproduce most of the exoplanets with the most favourable values of $\al=10^{-5}$ and $\vfrag=4$~m/s and that the low turbulence strength is the key parameter determining the success of pebble accretion scenario. By investigating the properties of exoplanets that are toughest to reproduce with the pebble accretion scenario, we found that the success of the model is determined by sufficiently massive ($\gtrsim 0.001~M_\oplus$) planetary core forming quickly enough ($\lesssim 2\cdot10^5$~years). In our model, this is determined by the complicated interplay of the disc mass, its metallicity, and the orbital distance of the planet. Two key parameters for the model's success are the stellar mass and orbital distance. The planets that fail to reach pebble isolation mass are those that have low classical isolation mass. We also found that although a higher metallicity of the parent star helps planet growth, the effect of metallicity is not as strong as those of the parent star mass or the orbital separation.

 Some future directions this model can take is to create an in-depth analysis of the individual planets that do not grow with the pebble accretion model. Further, more parameters can be analyzed through this approach instead of just $\al$, $\vfrag$ and stellar metallicity. A thorough parameter analysis can be performed to find more parameters (planetary,  environmental or stellar) that can influence pebble accretion. Through such a vast parameter study, we can also aim to understand the growth of those planets that the classical planetesimal model does not explain. Furthermore, pebble accretion predicts that a significant amount of outer disc material is delivered to the inner planets. From the studies of its isotopic composition, we know this is not necessarily true for Earth \citep[see, e.g.,][]{2021SciA....7.7601B}. Future exoplanet atmosphere studies will also be useful to constrain the formation models and confirm whether the close-in planet growth by pebble accretion is common.

\section*{Acknowledgements}

We thank the anonymous referee for the comments that helped us improve the paper's clarity. This research was possible thanks to Jayashree Narayan's stay at the Max Planck Institute for Solar System Research funded by the \href{https://www.daad.de/en/study-and-research-in-germany/scholarships/daad-scholarships/}{German Academic Exchange Service (DAAD)}. Joanna Dr{\k{a}}{\.z}kowska and Vignesh Vaikundaraman were funded by the European Union under the European Union’s Horizon Europe Research \& Innovation Programme 101040037 (PLANETOIDS). However, the views and opinions expressed are those of the authors only and do not necessarily reflect those of the European Union or the European Research Council. Neither the European Union nor the granting authority can be held responsible. 

\section*{Data Availability}
 
The data underlying this article will be shared on reasonable request to the corresponding author.


\bibliographystyle{mnras}
\bibliography{references} 

\bsp	
\label{lastpage}
\end{document}